%% file: ratiom_fB.tex
\documentclass[11pt]{article}
\usepackage{graphicx}
\usepackage[tight,TABTOPCAP]{subfigure}
\usepackage{amsmath}
\usepackage{amssymb}
\usepackage{cite}

\newcommand\pubnumber{~}
\newcommand\pubdate{\today}

\def\ode{$^a$CP$^{\,3}$-Origins \& Danish IAS, University of Southern
  Denmark,
  Odense M, Denmark}

\def\irtre{$^b$INFN, Sezione di Roma Tre,
  Rome, Italy}

\def\tov{$^c$Dipartimento di Fisica, Universit\`a di Roma ``Tor
  Vergata",\\
  INFN, Sezione di Tor Vergata,
  Rome, Italy}

\def\cf{$^d$Museo Storico della Fisica e Centro Studi e Ricerche ``Enrico
  Fermi",
  Rome,
  Italy}

\def\val{$^e$Departament de F\'{\i}sica Te\`orica and IFIC, Univ. de
  Val\`encia-CSIC,
  Val\`encia, Spain}

\def\mad{$^f$Instituto de F\'isica Te\'orica UAM/CSIC and Departamento
  de F\'isica Te\'orica,\\ Universidad Aut\'onoma de Madrid,
  Madrid, Spain}

\def\rtre{$^g$Dipartimento di Matematica e Fisica, Universit\`a degli
  Studi Roma Tre, Rome, Italy}

\def\liv{$^h$Theoretical Physics Division,
  The University of Liverpool, Liverpool,
  UK}

\def\soto{$^i$School of Physics and Astronomy, University of Southampton,
  Southampton, UK}

\def\jul{$^j$IAS, IKP, JCHP and JARA-HPC, Forschungszentrum J\"ulich,
  J\"ulich, Germany}

\def\Title#1{\begin{center} {\Large #1 } \end{center}}
\def\Author#1{\begin{center}{ \sc #1} \end{center}}
\def\Address#1{\begin{center}{ \it #1} \end{center}}

\newcommand\pubblock{\rightline{\begin{tabular}{l} \pubnumber\\
      \pubdate  \end{tabular}}}
\newenvironment{Abstract}{\begin{quotation}  }{\end{quotation}}
\newenvironment{Presented}{\begin{quotation} \begin{center} 
      PRESENTED AT\end{center}\bigskip 
      \begin{center}\begin{large}}{\end{large}\end{center} \end{quotation}}
\def\Acknowledgements{\bigskip  \bigskip \begin{center} \begin{large}
      \bf ACKNOWLEDGEMENTS \end{large}\end{center}}
\input econfmacros.tex
\newcommand{\nft}{N_\mathrm{f}=2}
\newcommand{\nftp}{N_\mathrm{f}=2+1+1}
\newcommand{\msbar}{\overline{\mathrm{MS}}}

\graphicspath{{figures/}}

\begin{document}
\begin{titlepage}
  \pubblock

  \vfill \Title{Lattice QCD Study of $B$-meson Decay Constants from ETMC}

  \vfill \Author{A.~Bussone\,$^a$, N.~Carrasco\,$^b$,
    P.~Dimopoulos\,$^{c,d}$, R.~Frezzotti\,$^c$, V.~Gim\'enez\,$^e$,
    G.~Herdo\'iza\,$^f$, P.~Lami\,$^{b,g}$, V.~Lubicz\,$^{b,g}$, C.~Michael\,$^h$,
    E.~Picca\,$^{b,g}$, L.~Riggio\,$^b$, G.C.~Rossi\,$^{c,d}$,
    F.~Sanfilippo\,$^i$, A.~Shindler\,$^j$,
    S.~Simula\,$^{b}$, C.~Tarantino\,$^{b,g}$\\
    [2mm]
    ETM Collaboration }

  \Address{\ode}    %%a
  \Address{\irtre}  %%b
  \Address{\tov}    %%c
  \Address{\cf}     %%d
  \Address{\val}    %%e
  \Address{\mad}    %%f
  \Address{\rtre}   %%g
  \Address{\liv}    %%h
  \Address{\soto}   %%i
  \Address{\jul}    %%j

  \vfill

%%%%%%%%%%%%%%%%%%%%%%%%%%%%%%%%%%%%%%%%%%%%%%%%%%%%%%%%%%%%%%%%%%%%%%%

  \begin{Abstract}

    We discuss a lattice QCD computation of the $B$-meson decay
    constants by the ETM collaboration where suitable ratios allow to
    reach the bottom quark sector by combining simulations around the
    charm-quark mass with an exactly known static limit. The different
    steps involved in this {\it ratio method} are discussed together
    with an account of the assessment of various systematic effects.
    A comparison of results from simulations with two and four flavour
    dynamical quarks is presented.

  \end{Abstract}

%%%%%%%%%%%%%%%%%%%%%%%%%%%%%%%%%%%%%%%%%%%%%%%%%%%%%%%%%%%%%%%%%%%%%%%

  \vfill

  \begin{Presented}
    8th International Workshop on the CKM \\ Unitarity Triangle (CKM 2014)\\
    Vienna, Austria, September 8-12, 2014
  \end{Presented}

  \vfill

\end{titlepage}

\def\thefootnote{\fnsymbol{footnote}}
\setcounter{footnote}{0}
%

%%%%%%%%%%%%%%%%%%%%%%%%%%%%%%%%%%%%%%%%%%%%%%%%%%%%%%%%%%%%%%%%%%%%%%%

\section{Introduction}

Flavour physics is the place of election where probing the scale of
possible extensions of the Standard Model (SM) can be carried out up
to energies considerably larger than the ones that can be attained in
present-day colliders such as the LHC. A detailed comparison of
experimental results to theoretical expectations based solely on the
SM is essential to determine its fundamental parameters. The study of
the consistency of the determinations of SM parameters by using as
many independent processes as possible provides a very stringent test
of the theory in its present formulation. For this program to be
successful non-perturbative QCD effects contributing to flavour
physics processes must be accurately evaluated. In particular lattice
QCD studies of hadronic matrix elements of weak operators are
essential to test unitarity constraints in the first two rows of the
CKM matrix.

Leptonic decays of $B$-mesons, $B \to \tau \nu$, have been thoroughly
studied at the $B$-factories. A reduction in the relative error on the
branching fraction, from $\sim20\%$ to $\sim5\%$, is expected to be
reached by Belle-II. An improvement in the extraction of the CKM
matrix element $|V_{\rm ub}|$ would thus also profit from a more
accurate lattice calculation of the $B$-meson decay constant,
$f_B$. Similarly, rare leptonic decays of neutral $B$-mesons are being
studied by LHC experiments. CMS and LHCb have measured the branching
fraction, ${\cal B}(B_s \rightarrow \mu^{+}\mu^{-})$, with a $\sim
25\%$ relative precision. In this case, the lattice determinations of
the $B_s$-meson decay constant, $f_{B_s}$, is an important ingredient
in the SM prediction of this branching fraction.

The study of the $b$-quark sector on the lattice requires the
development of a dedicated approach. The reason is that the $b$-quark
mass, $m_b \sim4.2\,{\rm GeV}$, is larger than the range of values of
the UV cutoff -- given by the inverse of the lattice spacing $a$ --
that can currently be probed in large-volume lattice QCD simulations,
$a^{-1} \lesssim 4\,{\rm GeV}$. The condition, $am_b \ll 1$, needed to
guarantee the convergence of an expansion in powers of the lattice
spacing is therefore not fulfilled. This leads to uncontrolled
systematic effects due to lattice artefacts. Several methods, relying
to some extent on the use of effective theories, have been developed
to address these difficulties (see Ref.~\cite{Aoki:2013ldr} for a
recent review of the various approaches).

In the following we will discuss the results of the determination of
the $B$-meson decay constants with a method that is based on the
construction -- directly in QCD -- of {\it ratios} of observables that
are well behaved from the point of view of lattice artefacts and that
have a precisely known static limit.

%%%%%%%%%%%%%%%%%%%%%%%%%%%%%%%%%%%%%%%%%%%%%%%%%%%%%%%%%%%%%%%%%%%%%%%

\section{Approaching the $b$-quark sector through the 
  ratio method}

To illustrate the {\it ratio method}~\cite{Blossier:2009hg}, let us
consider an observable, $\Xi(m_h)$, depending on the heavy quark mass
$m_h$ according to the scaling laws of heavy quark effective theory
(HQET),
\begin{equation}
  {\Xi(m_h)} \ = \ \Xi_{\rm stat} \ + \ \frac{d_{1}}{m_h} \ + \ {\rm O}\left(\frac{1}{m_h^2}\right)\,,
  \label{eq:xihqet}
\end{equation}    
where $\Xi_{\rm stat}$ refers to the observable in the static limit,
$m_h \to \infty$. A well known example is,
$\Xi(m_h)=\Phi_{h\ell}=f_{h\ell} \sqrt{M_{h\ell}}$, where $f_{h\ell}$
and $M_{h\ell}$ are the heavy-light ($h\ell$) pseudoscalar meson decay
constant and mass, respectively, depending on $m_h$. By considering a
geometric series of quark masses,
\begin{equation}
  m_h^{(i+1)} = \lambda\, m_h^{(i)}\,, \qquad i=1,2,3,\dots
  \label{eq:mhi}
\end{equation}    
with a constant step, $\lambda \gtrsim 1$, and an initial condition,
$m_h^{(1)} \approx m_c$, we can construct the following chain equation
for $\Xi$,
\begin{eqnarray}
  {\Xi(m_b) \ \equiv \ \Xi(m_h^{(K_b)})}  &=&
  {\Xi(m_h^{(1)})} \ \times
  \ {\frac{\Xi(m_h^{(2)})}{\Xi(m_h^{(1)})}}
  \ \times \ 
  \dots \ \times
  \  {\frac{\Xi(m_h^{(K_b)})}{\Xi(m_h^{(K_b-1)})}}\,,\\
  &=& {\Xi(m_h^{(1)})} \ \times \  \prod\limits_{i=2}^{K_b} R(m_h^{(i)})\,,
  \label{eq:chain}
\end{eqnarray}    
where the ratio $R(m_h^{(i)})$ is defined as follows,
\begin{equation}
  {R(m_h^{(i)})} \ \equiv \ \frac{\Xi\left(m_h^{(i)}\right)}{\Xi\left(m_h^{(i-1)}\right)}
  \,.
  \label{eq:rat}
\end{equation}    
In the example of eq.~(\ref{eq:chain}), we have assumed that $\lambda$
has been tuned in order to reach the $b$-quark sector after $K_b$
steps. As we will see, this can be achieved by setting $\lambda
\approx 1.18$ and $K_b=10$ but the value of $\lambda$ and the number
of steps $K_b$ can be adapted to the availability of lattice data and
to the target accuracy.

On a lattice regularisation of QCD that is improved to remove the
lattice artefacts of O$(a)$, the leading discretisation effects on a
quantity $\Xi(m_h)$ involving heavy quarks are of O$((am_h)^2)$. An
advantage of considering the ratio $R(m_h^{(i)})$ in
eq.~(\ref{eq:rat}) is that those cutoff effects become of order
$(\lambda-1)(am_h)^2$ and are therefore significantly suppressed for
$\lambda \gtrsim 1$.  We stress that the continuum limit can be taken
individually for each of the ratios $R(m_h^{(i)})$. The use of ratios
allows to control the continuum limit extrapolation up values of $m_h$
that are higher than those attainable by the observable $\Xi(m_h)$
alone. On currently accessible lattice ensembles, we observe that
values of $m_h \approx m_b/2$ can be reached, thus significantly
reducing the gap towards the $b$-quark mass.

From eq.~(\ref{eq:xihqet}), one expects that the $m_h$ dependence of
the ratios $R(m_h^{(i)})$ follows,
\begin{equation}
  R(m_h) \ = \ 1 \ + \ \frac{(\lambda-1)\,\hat{d}_{1}}{m_h} + {\rm
    O}\left(\frac{1}{m_h^2}\right)\,.
  \label{eq:rhqet}
\end{equation}    
The fact that $R(m_h)$ has a precisely known value in the static limit
allows to transform the approach to the $b$-quark sector into an {\it
  interpolation} that is constrained by direct determinations up to
$m_h \approx m_b/2$. It is also interesting to note that since the
$1/m_h$ term in eq.~(\ref{eq:rhqet}) is suppressed by the factor
$(\lambda-1) \approx 1/5$, it is in general expected that $1/m_h^2$ can also have
a sizeable contribution to the ratios $R(m_h)$.

The implementation of the {\it ratio method} can be summarised by the
following four steps:
\begin{itemize}

\item[(i)] Determination of the {\it triggering point},
  $\Xi(m_h^{(1)})$ in the r.h.s of eq.~(\ref{eq:chain}), through a
  direct calculation in the charm quark sector, $m_h^{(1)} \approx
  m_c$.

\item[(ii)] Calculation of the ratios $R(m_h^{(i)})$ for a
  $i=2,3,\dots, \bar{n}$, up to a mass $m_h^{(\bar{n})}$ that permits
  a proper control of the continuum-limit extrapolation of the ratio.

\item[(iii)] The previously computed ratios, $R(m_h^{(i)})$, are used
  to reach the the $b$-quark sector through an interpolation using the
  expected scaling laws of HQET and the fact that the static limit is
  known, see eq.~(\ref{eq:rhqet}).

\item[(iv)] The last step is to apply the chain equation
  eq.~(\ref{eq:chain}) to determine $\Xi(m_b)$.

\end{itemize}

The fact that the continuum limit is taken at each step of the {\it
  ratio method}, implies that it is a regularisation independent
approach that can be used with any lattice QCD regularisation. The
calculations are performed directly in QCD since the use of HQET only
intervenes in guiding the interpolation to the $b$-quark
sector\,\footnote{The knowledge of both the matching to QCD and the
  running of HQET operators is used to improve this interpolation
  step. Perturbative or also non-perturbative HQET results can
  therefore be incorporated in the method. We refer
  to Ref.~\cite{Carrasco:2013zta} for a discussion of the mixing of
  four-fermion operators.}.

Another attractive aspect of the {\it ratio method} is that its
computational cost is moderate since the computation of quark
propagators -- for the set of heavy quark masses -- can be performed
with a multi-mass inverter at the price of a single inversion in the
charm sector.
\begin{figure}[t!]
  %%\hspace*{-8.0cm}    
  \centering
  \subfigure[\label{fig:landa}]{
    \includegraphics[width=0.40\textwidth]{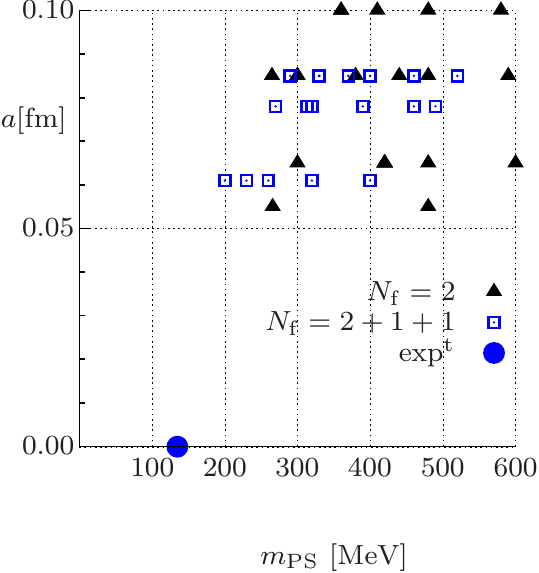}}
  \qquad
  \begin{minipage}[t]{0.46\textwidth} 
    \vspace*{-6.4cm}
    \subfigure[\label{fig:landb}]{
      \includegraphics[width=1.0\textwidth]{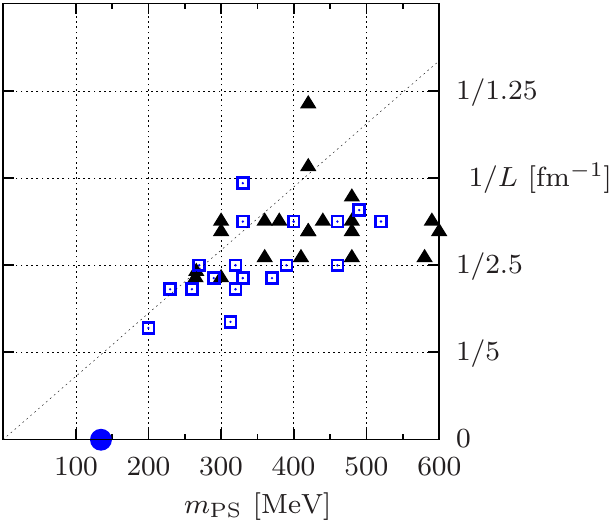}}
  \end{minipage}
  \caption{\small Range of values of the lattice spacing $a$, the
    simulated pion masses $M_{\rm PS}$ and the lattice size $L$, of
    ETMC ensembles with $\nft$ and $\nftp$ dynamical quark
    flavours. The physical pion point is denoted by the blue filled
    circle. The availability of various values of the lattice spacing
    is essential for a proper control of the continuum-limit
    extrapolation in the heavy-quark sector. The oblique line in the
    right panel denotes, $M_{\rm PS}L=3.5$, and the ensembles above this
    line are used to study finite volume effects.}
  \label{fig:land}
\end{figure}
%%

%%%%%%%%%%%%%%%%%%%%%%%%%%%%%%%%%%%%%%%%%%%%%%%%%%%%%%%%%%%%%%%%%%%%%%%

\section{Determination of $B$-meson decay constants}

\begin{figure}[t!]
  \centering
  \subfigure[\label{fig:ratfBa}]{
    \includegraphics[width=0.45\linewidth]{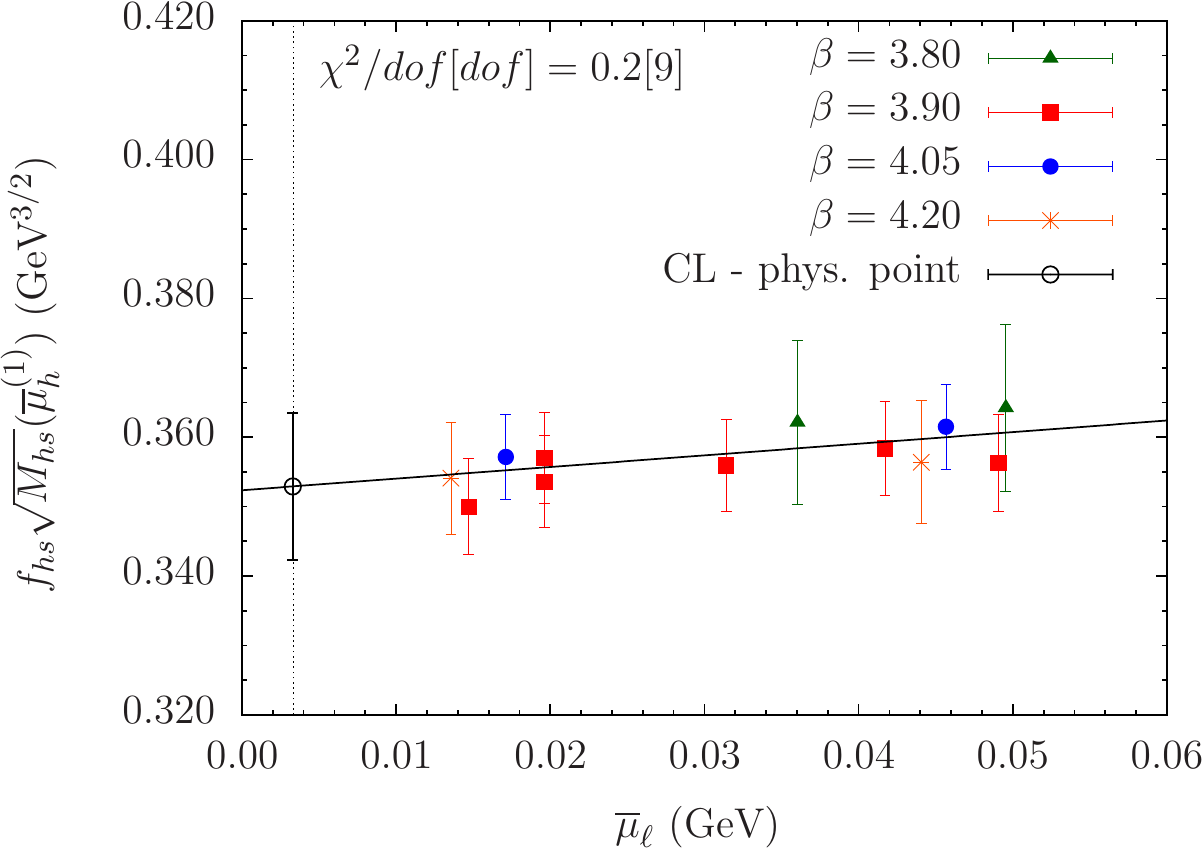}}
  \qquad
  \subfigure[\label{fig:ratfBb}]{
    \includegraphics[width=0.45\linewidth]{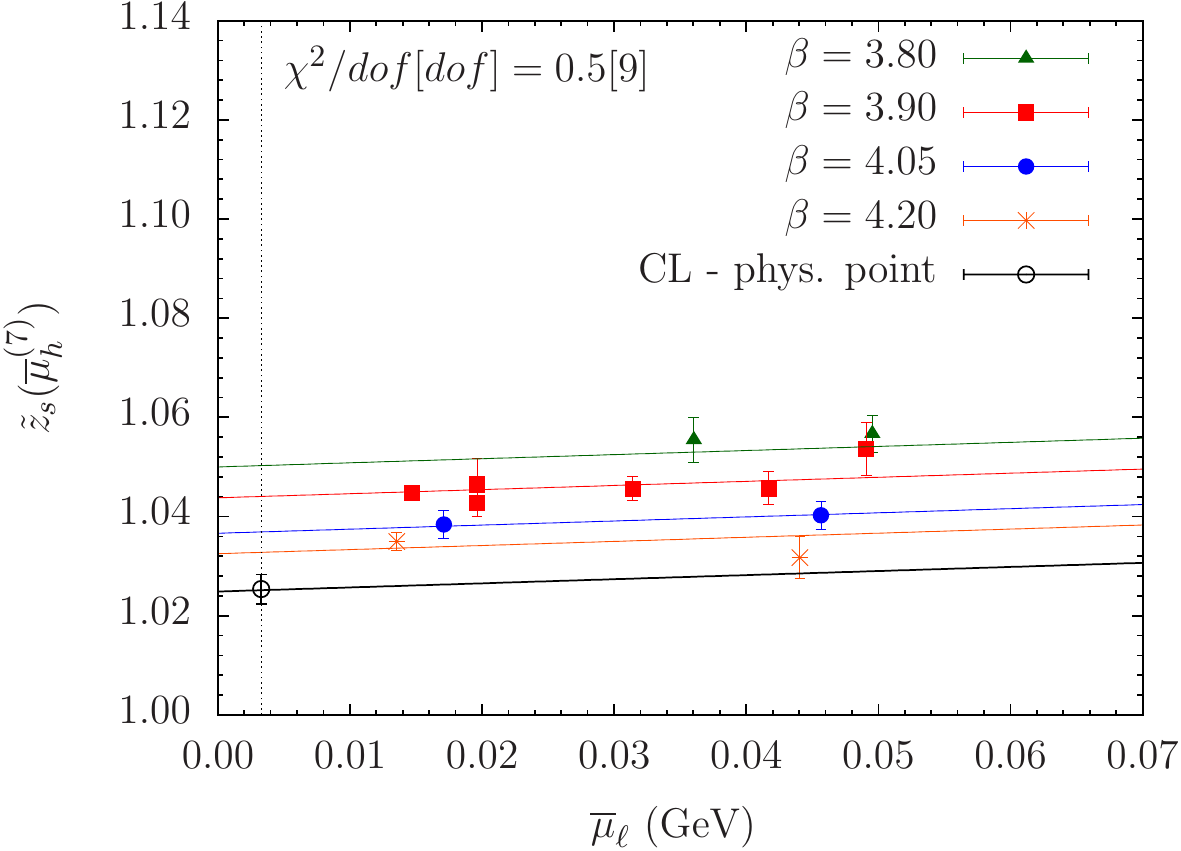}}
  \caption{\small (a) Results from $\nft$ dynamical simulations for
    the light-quark mass dependence of
    ${f_{hs}(\overline\mu_h^{(1)})}\,
    \sqrt{M_{hs}(\overline\mu_h^{(1)})}$ at the {\it triggering
      point}, corresponding to the charm quark,
    $\overline\mu_h^{(1)}=\overline\mu_c$. The good compatibility
    among measurements at four values of the lattice spacing (denoted
    by the values of the inverse bare coupling $\beta$ in the legend)
    indicates that this observable is not affected by large lattice
    artefacts. The result in the continuum limit (CL) and at the
    physical point is shown by the black empty
    circle~\cite{Carrasco:2013zta}. (b) Similarly, we illustrate the
    case of the heavier quark mass, with $\bar{n}=7$, for which the
    ratio $\tilde z_s(\overline\mu_h^{(7)})$ is computed. We observe
    that the use of a ratio allows to control the continuum limit
    extrapolation up to heavy-quark masses, $\overline\mu_h^{(7)}
    \approx 0.6 \,\overline\mu_b$.}
  \label{fig:ratfB}
\end{figure}
This study is based on lattice QCD simulations performed by the ETM
collaboration with $\nft$~\cite{Boucaud:2007uk,Boucaud:2008xu} and
$\nftp$~\cite{Baron:2010bv,Baron:2010th} flavours of Wilson twisted
mass fermions at maximal
twist~\cite{Frezzotti:2000nk,Frezzotti:2003ni}. Since the quark mass
is given by the twisted mass parameter $\mu$, in the following, we
will denote the heavy quark mass by, $\overline \mu_h = \overline
m_h$, where the $\msbar$ scheme at the scale $3\,{ \rm GeV}$ is
adopted. The set of available ensembles is illustrated in
Fig.~\ref{fig:land}, where the range of values of the lattice spacing
$a$, the simulated pion masses $M_{\rm PS}$ and the lattice size $L$
are shown. A first account of simulations at the physical pion mass
have been reported in Ref.~\cite{Abdel-Rehim:2013yaa}. ETMC has a
dedicated project aiming at precise determinations of observables
relevant for flavour physics involving
light~\cite{Blossier:2007vv,Blossier:2009bx,Baron:2009wt,Blossier:2010cr,Farchioni:2010tb,Baron:2011sf}
and
heavy~\cite{Blossier:2009hg,Dimopoulos:2011gx,Carrasco:2013zta,Carrasco:2013naa,Carrasco:2014cwa,Bussone:2014cha}
quarks.

The decay constant $f_{B_s}$ can be determined via the {\it ratio
  method} with various possible choices for the observable
$\Xi(\overline \mu_h)$. As an illustration we consider, $\Xi(\overline
\mu_h)=f_{hs} \sqrt{M_{hs}}/C_A^{\rm stat}(\overline\mu_h)$, but a
detailed analysis of how other choices are used assess systematic
effects was reported in Ref.~\cite{Carrasco:2013zta}. The factor
$C_A^{\rm stat}(\overline \mu_h)$ controls the matching between QCD
and HQET and the running of the axial current in the effective
theory. With this choice of $\Xi(\overline \mu_h)$, the chain equation
in eq.~(\ref{eq:chain}) can be written as follows,
\begin{equation}
  \tilde z_s(\overline\mu_h^{(2)})\ \times \ \tilde
  z_s(\overline\mu_h^{(3)})\ \times \ \ldots \ \times \ \tilde
  z_s(\overline\mu_h^{(10)}) \ = \ \frac{
    {f_{hs}(\overline\mu_h^{(10)})}\,
    \sqrt{M_{hs}(\overline\mu_h^{(10)})}}
  {{f_{hs}(\overline\mu_h^{(1)})}\,
    \sqrt{M_{hs}(\overline\mu_h^{(1)})}} \cdot \Big[ \frac{C^{\rm
        stat}_A(\overline\mu_h^{(1)})}{C^{\rm stat}_A(
      \overline\mu_h^{(10)})} \Big] \,,
  \label{eq:ratzs}
\end{equation}    
where $\tilde z_s(\overline\mu_h^{(i)})$ corresponds to the ratio in
eq.~(\ref{eq:rat}) for the specific choice of $\Xi(\overline
\mu_h)$. The {\it triggering point} has been chosen to precisely
correspond to the charm quark, $\overline\mu_h^{(1)}=\overline\mu_c$,
and the values, $\lambda=1.1784$ and $K_b=10$, have been tuned in
order to reach, $\overline\mu_h^{(K_b)}=\overline\mu_b$, by imposing
the physical value of the $B$-meson mass as an
input~\cite{Carrasco:2013zta}.

\begin{figure}[t!]
  \centering
  \subfigure[\label{fig:ratzRa}]{
    \includegraphics[width=0.45\linewidth]{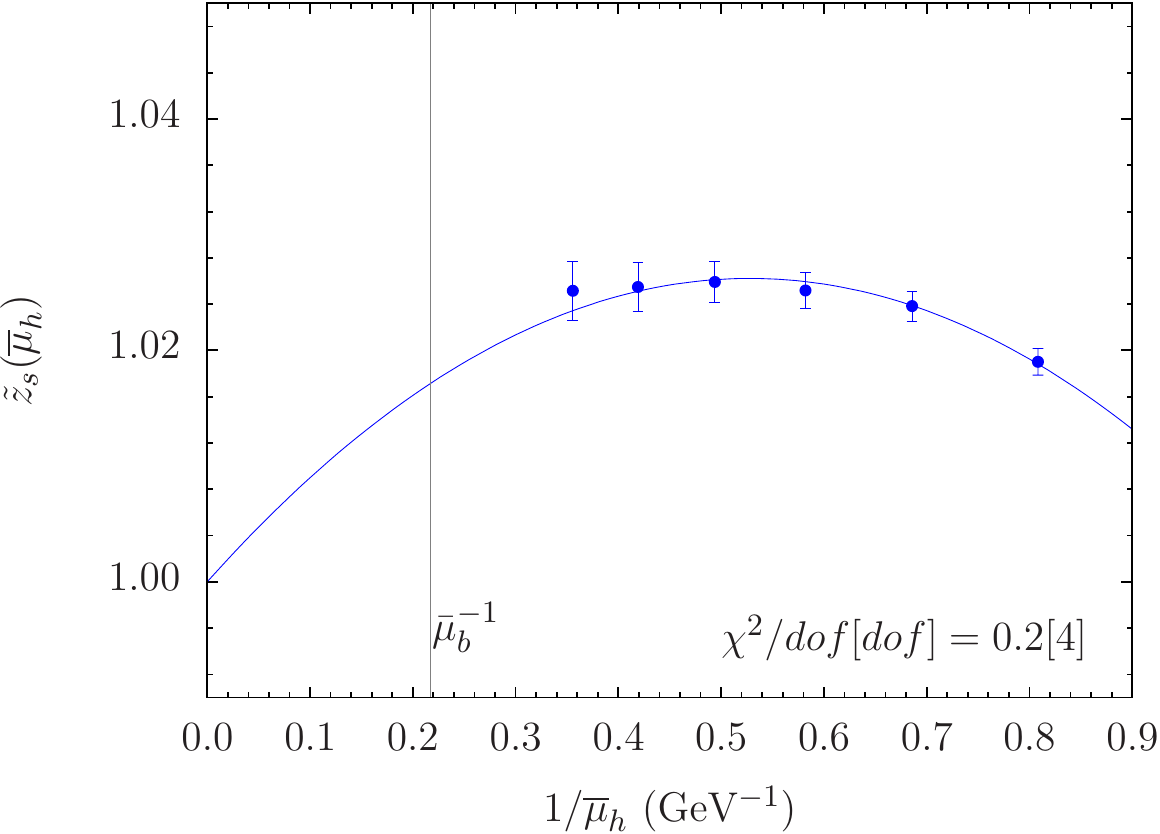}}
  \qquad
  \subfigure[\label{fig:ratzRb}]{
    \includegraphics[width=0.45\linewidth]{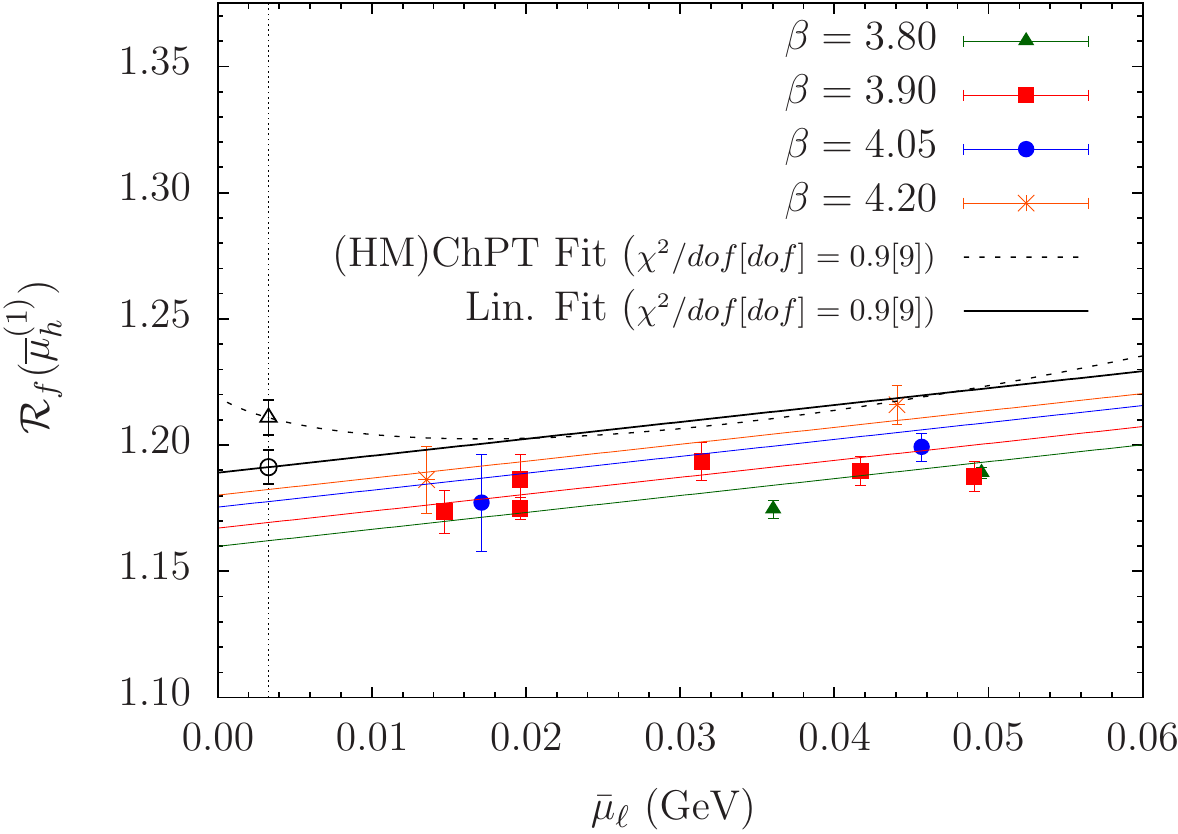}}
  \caption{(a) Dependence on the inverse of the heavy quark mass of
    the ratio $z_s(\overline\mu_h)$, in the continuum and for the
    light-quark physical point, measured in the interval $m_h \in
    [m_c; 0.6 m_b]$. The blue curve shows a quadratic fit inspired by
    the HQET expansion which is constrained by the precisely known
    value in the static limit. The $b$-quark mass, denoted by the
    vertical line, is thus reached by an interpolation. A detailed
    study of the assessment of possible systematic effects was
    reported in Ref.~\cite{Carrasco:2013zta}. (b) Light-quark mass
    dependence of the double ratio ${\cal R}_f$ in eq.~(\ref{eq:Rf})
    -- used to determine $f_{B_s}/f_B$ -- at the {\it triggering
      point}. The deviation between the results of chiral
    extrapolations based on a linear fit and on a heavy-meson chiral
    perturbation theory (HMChPT) ansatz is included in the systematic
    error analysis.}
  \label{fig:ratzR}
\end{figure}

The first step of the {\it ratio method} is illustrated in
Fig.~\ref{fig:ratfBa} where the continuum and chiral extrapolations of
$f_{hs} \sqrt{M_{hs}}$ at the {\it triggering point} is shown. The ratios
$\tilde z_s(\overline\mu_h^{(i)})$ are then extracted for
$i=2,3,\dots, \bar{n}$. Fig.~\ref{fig:ratfBb} shows that lattice
artefacts are under control up to $\bar{n}=7$, corresponding to the
heaviest mass, $\overline\mu_h^{(7)} \approx 0.6
\,\overline\mu_b$. The interpolation of the ratios $\tilde
z_s(\overline\mu_h^{(i)})$ to the $b$-quark sector is presented in
Fig.~\ref{fig:ratzRa}. Based on eq.~(\ref{eq:rhqet}), a quadratic fit
ansatz inspired by the HQET expansion is constrained to be equal to
$1$ in the static limit. The last step of the {\it ratio method} is to
use the chain equation in eq.~(\ref{eq:ratzs}) to determine $f_{B_s}
\equiv f_{hs}(\overline\mu_h^{(10)})$.

The ratio $f_{B_s}/f_B$ can be determined with an observable,
$\Xi(\overline\mu_h) = {\cal R}_f(\overline\mu_h)$, defined through a
double ratio, in the following way,
\begin{equation}
  {\cal R}_f(\overline\mu_h) \ =
  \ \left[\,\frac{\left(\frac{f_{hs}(\overline\mu_h)}{f_{hl}(\overline\mu_h)}\right)}{\left(
      \frac{f_{sl}}{f_{ll}} \right)}\,\right]\ \times\ \left(\frac{f_K}{f_\pi}\right)\,,
  \label{eq:Rf} 
\end{equation}
where $f_K$ and $f_\pi$ refer to the physical values of the kaon and
pion decay constants, respectively. The chiral extrapolation of ${\cal
  R}_f(\overline\mu_h)$ at the {\it triggering point} is shown in
Fig.~\ref{fig:ratzRb}. The successive steps of the {\it ratio method}
applied to ${\cal R}_f(\overline\mu_h)$ are not significantly
different than those shown in Figs.~\ref{fig:ratfB} and
\ref{fig:ratzRa} and can be found in Ref.~\cite{Carrasco:2013zta}.

\begin{table}[!t]
  \begin{center}
    \begin{tabular}{lccc}
      \hline 
      source of uncertainty $[\%]$ & $f_{B_s}$ & $f_{B_s}/f_{B}$ & $f_{B}$ \tabularnewline
      \hline
      stat. + fit~~(C.L. and chiral) & 2.2 & 0.8 & 2.1 \tabularnewline
      \hline 
      lat. scale & {2.0} & - & {2.0} \tabularnewline
      \hline 
      discr. effects & 1.3 & 0.4 & 1.7 \tabularnewline
      \hline 
      $1/\mu_{h}$ & 1.0 & 0.1 & 1.1 \tabularnewline
      \hline 
      chiral extr. trig. point & - & {1.7} & 1.7 \tabularnewline
      \hline 
      {total}  & {3.4} & {2.0} & {4.0} \tabularnewline
      \hline
    \end{tabular}
  \end{center}
  \caption{Relative contributions (in percent) to the uncertainties in
    the determination of $f_{B_s}$, $f_{B_s}/f_{B}$ and $f_{B}$ from
    our studies with $\nft$ sea quarks. An overall relative error of
    $2\%$ is achieved for the dimensionless ratio, $f_{B_s}/f_{B}$,
    while systematic effects coming in particular from scale setting
    lead to larger relative errors for the individual decay
    constants. We refer to Ref.~\cite{Carrasco:2013zta} for a complete
    description of the procedures used to determine the statistical
    and systematic uncertainties.}
  \label{tab:error}
\end{table}   

\begin{figure}[t!]
  \centering \subfigure[\label{fig:compfB}]{
    \includegraphics[width=0.45\linewidth]{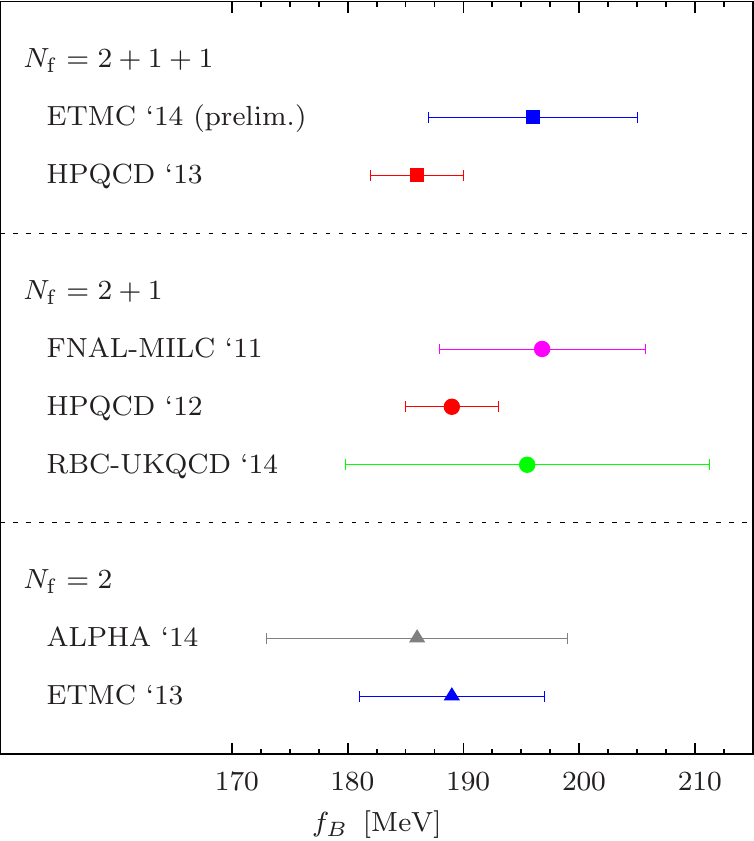}} \qquad
  \subfigure[\label{fig:compfBs}]{
    \includegraphics[width=0.45\linewidth]{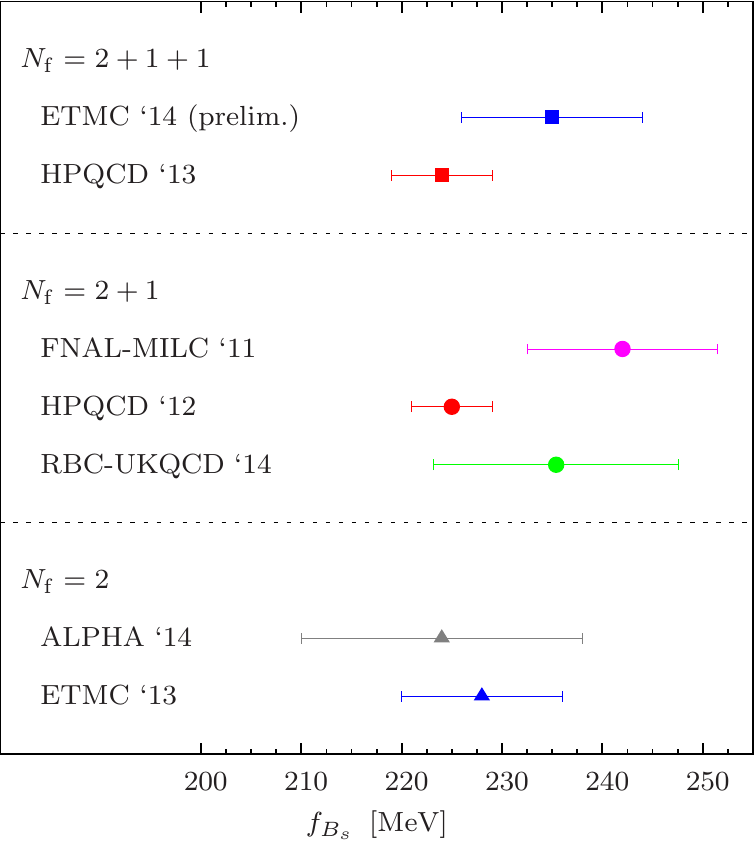}}
  \caption{Comparison of recent lattice QCD
    determinations~\cite{Carrasco:2013naa,Dowdall:2013tga,Bazavov:2011aa,Na:2012kp,Christ:2014uea,Bernardoni:2014fva,Carrasco:2013zta}
    of the decay constants $f_B$ and $f_{B_s}$ from simulations with
    $\nft$, $N_{\rm f}=2+1$ and $\nftp$ flavours of dynamical
    quarks. We refer to the recent reviews in
    Refs.~\cite{Aoki:2013ldr,Bouchard:lat14,ElKhadra:ckm14,Lucha:2014hqa,Tantalo:ckm14}
    for more detailed accounts of these computations.}
  \label{fig:compa}
\end{figure}

The results for the decay constants in the $\nft$ case read~\cite{Carrasco:2013zta},
\begin{align}
  f_{B_s} = 228(8)\,{\rm MeV}  \quad , \quad
  f_{B} = 189(8) \,{\rm MeV} \quad , \quad
  \dfrac{f_{B_s}}{f_B}= 1.206(24)\,.
  \label{eq:resnft}
\end{align}
The decomposition of the different contributions to the relative error
are given in Table~\ref{tab:error}. Preliminary results from $\nftp$
simulations~\cite{Carrasco:2013naa},
\begin{align}
  f_{B_s} = 235(9)\,{\rm MeV}  \quad , \quad
  f_{B} = 196(9) \,{\rm MeV} \quad , \quad
  \dfrac{f_{B_s}}{f_B}= 1.201(25)\,,
  \label{eq:resnftp}
\end{align}
are compatible with the $\nft$ determinations in
eq.~(\ref{eq:resnft}). This suggests that for these observables,
sea-quark effects from strange and charm quarks are smaller than the
present accuracy. A comparison of results from various recent lattice
QCD determinations is shown in Fig.~\ref{fig:compa}.

%% 

%%%%%%%%%%%%%%%%%%%%%%%%%%%%%%%%%%%%%%%%%%%%%%%%%%%%%%%%%%%%%%%%%%%%%%%

\section{Conclusions}

Physical processes in the quark-flavour sector of the Standard Model
can be used to put severe constraints on the parameter space of new
physics models. Leptonic decays of $B$-mesons are a remarkable example
of processes that are currently being constrained both from the
experimental and the theory side. Lattice QCD calculations of the
$B$-meson decay constants are needed to determine the SM predictions
of these decay rates. We have described the results of a method that
considers suitable ratios to constrain the approach to the $b$-quark
sector. A comparison of various lattice determinations indicates that
$f_B$ and $f_{B_s}$ can currently be determined with a few percent
accuracy. No significant effects from strange and charm sea quarks can
be observed at this level of precision.

The {\it ratio method} has also been used for the determination of
other observables in the bottom sector. These include the $b$-quark
mass~\cite{Blossier:2009hg,Dimopoulos:2011gx,Carrasco:2013zta,Carrasco:2013naa},
$B$-mixing bag parameters for the complete basis of four-fermion
operators~\cite{Carrasco:2013zta} or the form factors of semi-leptonic
$B$-decays~\cite{Atoui:2013zza,Sanfilippo:ckm14}. It is interesting to
note that preliminary studies with $\nftp$ dynamical quarks indicate
that an accuracy $\lesssim 2\%$ can be achieved for the ratio of quark
masses, $m_b/m_c$, via the {\it ratio method}\cite{Bussone:2014cha}.

%%%%%%%%%%%%%%%%%%%%%%%%%%%%%%%%%%%%%%%%%%%%%%%%%%%%%%%%%%%%%%%%%%%%%%%

\Acknowledgements

G.H. thanks the organisers for the stimulating environment created
during this event.
We are grateful to all members of ETMC for fruitful discussions. 
We acknowledge the CPU time provided by the PRACE Research
Infrastructure under the project PRA067 at the J\"ulich and CINECA
SuperComputing Centers, and by the agreement between INFN and CINECA
under the specific initiative INFN-lqcd123.
We acknowledge computer time made available to us on the Altix system
at the HLRN supercomputing service in Berlin under the project
``B-physics from lattice QCD simulations''.
G.H. acknowledges support by the Spanish MINECO through the Ram\'on y
Cajal Programme and through the project FPA2012-31686 and by the
Centro de Excelencia Severo Ochoa Program SEV-2012-0249.
%%

%%%%%%%%%%%%%%%%%%%%%%%%%%%%%%%%%%%%%%%%%%%%%%%%%%%%%%%%%%%%%%%%%%%%%%%

\end{document}

%% file: econfmacros.tex
\def\beq{\begin{equation}}
\def\eeq#1{\label{#1}\end{equation}}
\def\eeqn{\end{equation}}

\def\beqa{\begin{eqnarray}}
\def\eeqa#1{\label{#1}\end{eqnarray}}
\def\eeqan{\end{eqnarray}}

\let\bar=\overbar

\def\Dslash{\not{\hbox{\kern-4pt $D$}}}
\def\dslash{\not{\hbox{\kern-2pt $\del$}}}

\def\msb{{\bar{\ssstyle M \kern -1pt S}}}